\documentclass[fleqn,usenatbib]{mnras}

\usepackage{newtxtext,newtxmath,verbatim}

\usepackage[T1]{fontenc}
\usepackage{ae,aecompl}

\usepackage{graphicx}	
\usepackage{amsmath}	
\usepackage{amssymb}	

\newcommand{\adag}{g_{\dagger}}
\newcommand{\ffb}{ \langle \dot{p}/ m_{\ast} \rangle}
\newcommand{\fmassive}{f_{\rm massive}}

\newcommand{\gcrit}{g_{\rm crit}}
\newcommand{\tildegcrit}{\tilde{g}_{\rm crit}}

\title[The universal acceleration scale from stellar feedback]{The universal acceleration scale from stellar feedback}

\author[Grudi\'{c} et al.]{
Michael Y. Grudi\'{c}$^{1,3}$\thanks{E-mail: mike.grudic@northwestern.edu},
Michael Boylan-Kolchin$^{2}$,
Claude-Andr\'{e} Faucher-Gigu\`{e}re$^{1}$,\newauthor 
and Philip F. Hopkins$^{3}$ 
\\
$^{1}$Department of Physics and Astronomy and CIERA, Northwestern University, 2145 Sheridan Road, Evanston, IL 60208, USA\\
$^{2}$Department of Astronomy, The University of Texas at Austin, 2515 Speedway, Stop C1400, Austin, TX 78712, USA \\
$^{3}$TAPIR, Mailcode 350-17, California Institute of Technology, Pasadena, CA 91125, USA
}

\date{Accepted XXX. Received YYY; in original form ZZZ}

\pubyear{2020}
\begin{document}
\label{firstpage}
\pagerange{\pageref{firstpage}--\pageref{lastpage}}
\maketitle

\begin{abstract}
It has been established for decades that rotation curves deviate from the Newtonian gravity expectation given baryons alone below a characteristic acceleration scale $\adag\sim 10^{-8}\,\rm{cm\,s^{-2}}$, a scale promoted to a new fundamental constant in MOND. In recent years, theoretical and observational studies have shown that the star formation efficiency (SFE) of dense gas scales with surface density, SFE $\sim \Sigma/\Sigma_{\rm crit}$ with $\Sigma_{\rm crit} \sim \langle\dot{p}/m_{\ast}\rangle/(\pi\,G)\sim 1000\,\rm{M_{\odot}\,pc^{-2}}$ (where $\langle \dot{p}/m_{\ast}\rangle$ is the momentum flux output by stellar feedback per unit stellar mass in a young stellar population). We argue that the SFE, more generally, should scale with the local gravitational acceleration, i.e. that SFE $\sim g_{\rm tot}/\gcrit  \equiv (G\,M_{\rm tot}/R^{2}) / \langle\dot{p}/m_{\ast}\rangle$, where $M_{\rm tot}$ is the total gravitating mass and $\gcrit=\langle\dot{p}/m_{\ast}\rangle = \pi\,G\,\Sigma_{\rm crit} \approx 10^{-8}\,\rm{cm\,s^{-2}} \approx\adag$. Hence the observed $g_\dagger$ may correspond to the characteristic acceleration scale above which stellar feedback cannot prevent efficient star formation, and baryons will eventually come to dominate. We further show how this may give rise to the observed acceleration scaling $g_{\rm obs}\sim(g_{\rm baryon}\,\adag)^{1/2}$ (where $g_{\rm baryon}$ is the acceleration due to baryons alone) and flat rotation curves. The derived characteristic acceleration $\adag$ can be expressed in terms of fundamental constants (gravitational constant, proton mass, and Thomson cross section): $\adag\sim 0.1\,G\,m_{p}/\sigma_{\rm T}$.
\end{abstract}



\section{Introduction}
The kinematics of galaxies require something beyond the Newtonian gravity of baryonic matter \citep[e.g.,][]{rubin:1978.rotation.curves,rubin:1980.rotation.curves, bosma:1981.rotation.curves, bosma:1981.rotation.curves2}, an amazing observation that has moved from controversial to iron-clad over the past five decades \citep[e.g.,][]{bershady:2011.submaximal}. In conjunction with observations of the cosmic microwave background and large-scale structure of the Universe \citep[e.g.,][]{planck2018:parameters}, this observation has led to the commonly-accepted idea that the mass content of the Universe is dominated by dark matter. 
In the standard $\Lambda$ cold dark matter ($\Lambda$CDM) cosmology, dark matter reconciles observations with the fact that, according to general relativity, Newtonian gravity should be valid throughout the Universe whenever gravitational fields are weak.

The effects of dark matter become important not at some characteristic radius, nor some characteristic galaxy mass, but at a critical \textit{acceleration} scale. For gravitational accelerations $g \gg \adag \approx 1.2 \times 10^{-8}\,{\rm cm\,s^{-2}}$, baryons dominate the gravitational dynamics of galaxies, while for $g \ll \adag$, dark matter dominates. This led \citet{mond,mond2,mond3} to propose that modifying Newton's law of gravity (or inertia) for very small accelerations, rather than dark matter, is the correct interpretation of galaxy dynamics. More recently, \citet{lelli:2017.rar} 
have shown that the total gravitational acceleration in disk galaxies is well-predicted by the acceleration provided by the baryons alone, with the acceleration scale $\adag$ as the transition point from where the observed acceleration agrees with that of the baryons and below which some additional acceleration is required. 

Some have suggested that a characteristic acceleration scale is an expected outcome of galaxy formation in $\Lambda$CDM. \citet{kaplinghat:2002.lcdm.mond} argued that $\Lambda$CDM predicts galaxies to have an insensitive acceleration scale at the radius where the transition from baryon-dominated to DM-dominated occurs. 
However, Kaplinghat \& Turner's derivation assumed (incorrectly) that most of the available baryons collapse into galactic disks in all DM halos. 
This derivation therefore cannot account 
for low surface brightness galaxies that have acceleration $g < g_\dagger$ everywhere. 
\citet{vdb.dalcanton:analytic} showed that semianalytic galaxy models could be constructed within $\Lambda$CDM\ that exhibit the observed acceleration scale $g_\dagger$. They accounted for stellar feedback with a simple parametrized model, and found that these parameters could be tuned to put the model on observed galactic scaling relations. However it was not clear whether stellar feedback actually acts according to this model, and if so, how the value of $g_\dagger$ could be understood from basic principles or stellar evolution considerations. 

Most recently, many cosmological galaxy formation simulations with baryonic physics have formed galaxies with rotation curves in good agreement with the RAR \citep[e.g.,][]{santos:2016.rar,keller:2017.lcdm.rar, navarro:2017.rar,ludlow:2017.eagle.rar, garaldi:2018.rar,dutton:2019.nihao.rar}.  As in \citet{vdb.dalcanton:analytic}'s semianalytic model, these simulation studies generally find feedback to be crucial for forming galaxies with realistic properties in $\Lambda$CDM. In the context of $\Lambda$CDM, it is clear that feedback is what connects the baryonic content of a disk galaxy to its dark matter halo and that this connection must be in some sense ``universal'' to explain observed scaling relations. 
In this Letter, we argue that stellar feedback in the form of momentum injection from massive stars can naturally explain the critical acceleration scale found in observations of disk galaxies, with that scale {\em built into} stellar physics and evolution, as opposed to arising from a coincidence or ``fine-tuning'' of various effects.

\vspace{-0.5cm}
\section{Physical Model}

\begin{figure}
    \centering
    \includegraphics[width=0.4\textwidth]{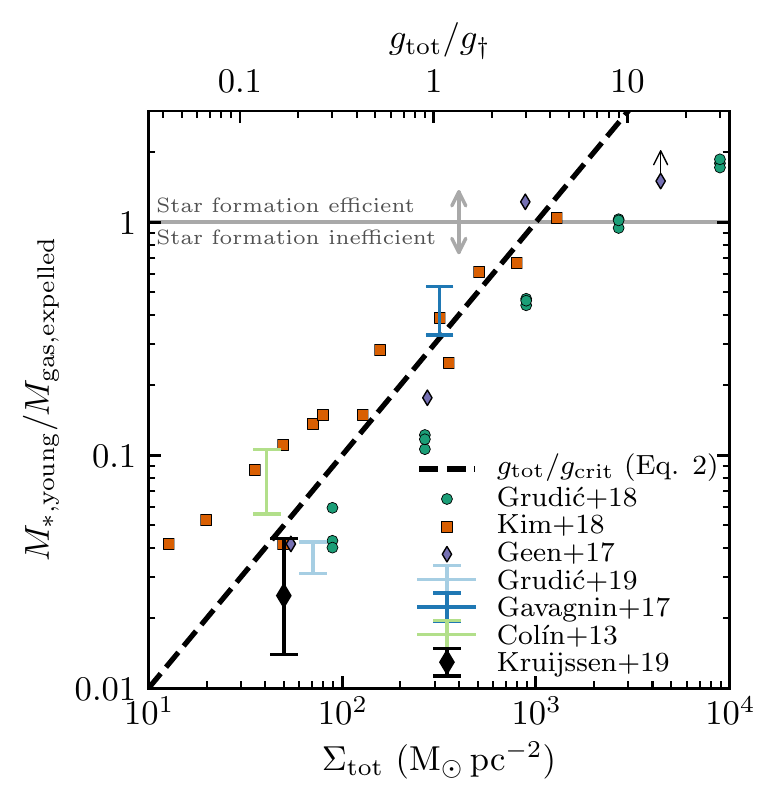}
    \caption{
   Ratio of the young stellar mass formed to the expelled gas mass in GMCs as a function of total mass surface density $\Sigma_{\rm tot}$, which is equivalent to an effective acceleration $g_{\rm tot} \equiv \pi G \Sigma_{\rm tot}$, plotted on top in units of $\adag$ (the observed characteristic acceleration scale of galaxies). \citet{kruijssen:2019.ngc300} is the one observational measurement presently available, and was made in NGC300; all other points are theoretical results. The models generally predict that $M_{\ast,\rm young}/M_{\rm expelled} \rightarrow 1$ as $\Sigma_{\rm tot}$ exceeds $\Sigma_{\rm crit} \sim 10^3 \,\rm M_\odot\,pc^{-2}$, i.e. star formation becomes efficient.
     The theoretical results come from (magneto)hydrodynamic simulations \citep{ colin:2013.rhd.sf.sims, geen:2017, gavagnin:2017.rhd.cluster.formation, grudic:2016.sfe, kim:2018.rhd, grudic:2018.gmc.sfe}, and all include stellar feedback at least in the form of radiation from massive stars. The dashed line shows the simple analytic scaling from Eq. \ref{eqn:sfe.agrav} for comparison. 
     Error bars denote the full range of results obtained in studies that survey only one value of $\Sigma_{\rm eff}$ and vary either cloud mass or physics prescriptions.
  }

    \label{fig:sfe}
\end{figure}


\subsection{Scaling of the Star Formation Efficiency}
\label{section:derivation}
Star formation efficiency (SFE), broadly speaking, is the mass in stars formed per mass in gas available for star formation. The SFE has numerous observational and theoretical definitions, depending on the spatial (e.g. galactic vs. $\mathrm{kpc}$-scale vs. cloud-scale) and temporal (e.g. per free-fall time, per Gyr, or integrated) scales of interest, as well as the observational tracer used (e.g. YSOs, $H\alpha$). 
We refer to \citet{krumholz:2018.star.cluster.review} for a review of the SFE on GMC/star cluster scales and to \citet{kennicutt:2012.review} for galactic scales. For the simple discussion here, we define the SFE as the fraction of available gas mass converted to new stars during a given ``episode" of star formation, $\mathrm{SFE}\equiv\frac{M_{\rm \ast,young}}{M_{\rm gas,initial}}$. The scaling of the SFE which results from a competition between gravity and stellar feedback has been extensively studied over the last few decades, both in the context of individual star-forming clouds and for entire galaxies \citep[e.g.][]{larson:1974.sne.fb,rees.ostriker:1977.sne.fb, dekel.silk:1986, silk:1997.fb.regulated.sf, efstathiou:2000.sne.fb, murray:2005.winds, murray:molcloud.disrupt.by.rad.pressure, fall:2010.sf.eff.vs.surfacedensity}. 
Below we sketch a toy model that captures the core of our present understanding of self-regulated star formation.

Consider a ``patch'' of gas in a galactic disk (or cloud) of mass $M_{\rm gas,initial}$ localized within a radius $R$ which is Jeans unstable and begins to cool and form stars. These stars will act back on the collapsing gas, injecting momentum at a rate per unit area $\dot{P}_{\rm fb}$ which is proportional to the mass of young stars (since the feedback is dominated by massive stars), $\dot{P}_{\rm fb} \sim \langle \dot{p}/{m}_{\ast} \rangle\, M_{\ast,\,{\rm young}}$ (where $\langle \dot{p}/{m}_{\ast} \rangle$ is the momentum injection rate per stellar mass formed). If this exceeds the weight of the gas $\sim G\,M_{\rm tot}\,M_{\rm gas}/R^{2}$ (where $M_{\rm tot}$ is the {\it total} mass localized within $<R$), then the gas becomes unbound, i.e.\ SF ceases and the remaining gas is ejected, when 
\begin{align}
\label{eqn:sfe}    \frac{M_{\ast,\,{\rm young}}}{M_{\rm gas,\,expelled}} &\sim \frac{\,G\,M_{\rm tot}}{\langle \dot{p}/{m}_{\ast} \rangle\,R^{2}} = \frac{\Sigma_{\rm tot}}{\Sigma_{\rm crit}}, %
\end{align}
where $\Sigma_{\rm tot}\equiv M_{\rm tot}/\uppi R^2$ and $\Sigma_{\rm crit}\equiv \langle \dot{p}/{m}_{\ast} \rangle / (\pi\,G)$. 

In recent years, a considerable body of work has explored the SFE and demonstrated that such a scaling, with a roughly constant $\Sigma_{\rm crit}\sim 1000\,\rm{M_{\odot}\,pc}^{-2}$, works remarkably well at describing both observations \citep[e.g.,][]{wong:2019.lmc.sf} and detailed numerical simulations of cloud collapse \citep[e.g.,][]{hopkins:fb.ism.prop, colin:2013, raskutti:2016.gmcs, gavagnin:2017.rhd.cluster.formation, grudic:2016.sfe, kim:2018.rhd, li:2019.cfe}. 
We compile in Figure \ref{fig:sfe} SFE predictions for GMCs (or more generally ``ISM patches") from various simulations including stellar feedback, as well as recent observations of NGC300 \citep{kruijssen:2019.ngc300}. The compilation shows that there is an emerging consensus that the SFE scales linearly with $\Sigma_{\rm tot}$ over at least $2~\mathrm{dex}$ in surface density, reaching $\sim 1$ for $\Sigma_{\rm tot} \gtrsim \Sigma_{\rm crit}$.\footnote{It is common in the star formation literature to define the SFE as $M_{\ast,\,{\rm young}}/(M_{\rm gas,\,expelled} + M_{\ast,\,{\rm young}})$, effectively 
neglecting the possibility of a significant dark matter component.} 
Although model predictions vary at the factor $\sim 2-3$ level at fixed $\Sigma_{\rm tot}$, much of the scatter can be attributed to differences in initial conditions, definitions, and numerical methods \citep[e.g.,][]{geen:2018.sf.chaos,hopkins.grudic:2018.rp,elephant}.

 Moreover, the ensemble and time-averaged version of this simple scaling can explain the observed Schmidt-Kennicutt relation in terms of $\langle \dot{p}/m_{\ast} \rangle$
 \citep[or its time-integrated-equivalent $\langle p/m_{\ast} \rangle$; e.g.,][]{silk:1997.fb.regulated.sf, thompson:rad.pressure, ostriker.shetty:2011,  hopkins:rad.pressure.sf.fb, cafg:sf.fb.reg.kslaw, semenov:2016, orr:2018.kennicutt.schmidt}.  Taken one step further, this also naturally leads to the scalings for momentum-conserving galactic outflows 
 \citep[e.g.,][]{murray:2005.winds, hopkins:stellar.fb.winds, hayward.hopkins:2017}, 
with the same constant $\langle \dot{p}/m_{\ast} \rangle$  appearing.\footnote{If instead of the non-equilibrium derivation of Eq.~(\ref{eqn:sfe}), one considers a time-steady galactic SFR $\dot{M}_{\ast}$ and wind mass loss $\dot{M}_{\rm out}$ with momentum flux $\dot{M}_{\rm out}\,v_{\rm escape} \sim \langle p/m_{\ast} \rangle \dot{M}_{\ast} \sim \ffb\,M_{\ast,\,\rm young}$ (where $\langle p/m_{\ast}\rangle = \int \ffb\,dt \sim t_{\ast} \ffb$ for a single stellar population), one obtains the usual momentum-driven wind scaling $\dot{M}_{\rm out} \propto 1/v_{\rm escape}$ \citep[e.g.,][]{murray:2005.winds, dave:2011.galactic.outflows} and we can write $M_{\ast} / M_{\rm gas,\,expelled} \sim \langle \dot{M}_{\ast} \rangle / \langle \dot{M}_{\rm out} \rangle \sim v_{\rm escape}/\langle p/m_{\ast} \rangle \sim (t_{\ast}\,\Omega)^{-1}\,(g_{\rm tot}/\gcrit)$, where $t_{\ast}\,\Omega\sim1$ is a correction applicable when $\Omega^{-1} \gtrsim t_{\ast}$.}

The value $\langle \dot{p}/m_{\ast} \rangle$ thus plays a critical and ``universal'' role in star formation.  It is important to note that the value of $\langle \dot{p}/m_{\ast} \rangle$ is similar regardless of whether radiation pressure, photoionization, or supernovae (SNe) dominate the feedback, as they all provide similar momentum injection rates for standard stellar population models \citep[e.g.,][]{starburst99, bruzual:2003.ssp, agertz:2013.new.stellar.fb.model, hopkins:rad.pressure.sf.fb}. Specifically, radiation pressure from a young stellar population contributes the specific momentum of photons, $\langle \dot{p}/m_{\ast} \rangle \approx c^{-1}\,(L/M)_{\ast,\,{\rm young}} \sim c^{-1}\,(1000\,L_{\odot}/M_{\odot}) \approx 6\times 10^{-8} \mathrm{cm\,s}^{-2}$. The simulations of \citet{colin:2013}, \citet{kim:2018.rhd} and \citet{geen:2017} predict a similar effective $\langle\dot{p}/m_{\ast}\rangle$ when including photoionization alone: this is generally understood to be because HII bubbles ``vent out'' their energy in inhomogeneous GMCs, so this feedback channel also behaves effectively as a momentum-driven. SNe from a given stellar population impart a momentum per unit stellar mass of $\sim1000\mathrm{km\,s}^{-1}$, nearly independent of ambient medium properties \citep[e.g.,][]{martizzi:2015.snr.inhomogeneous, kim:2015.snr.mom}, and within that population core-collapse SNe occur steadily until the last (i.e. least-massive, $\sim 8 M_\odot$) progenitors go off around $40\mathrm{Myr}$. This again yields a momentum injection rate of $\langle \dot{p}/m_{\ast} \rangle \approx {\rm 1000~km~s^{-1}} / 40~\mathrm{Myr} \approx 6 \times 10^{-8}~\mathrm{cm\,s^{-1}}$. Lastly, in standard stellar mass loss models (which likely overestimate the effects of winds; \citealt{smith:2014.massloss}), the momentum from stellar winds is at most comparable to that of photons.

Although myriad controversies and unsolved questions still remain in the study of star formation,  
 this controversy has largely centred on the exact order-unity coefficients (e.g. the fraction of the momentum coupled vs.\ ``vented,'' corrections for turbulent media, cooling and chemistry effects 
 [\citealt{clark.glover:2014}], and non-linear time-dependent effects), and the question of which feedback mechanisms dominate on which spatial and timescales. 
 None of these effects are expected to alter the dimensional scaling in Eq.~\ref{eqn:sfe} at the level of our simplified explanation. 
\vspace{-0.4cm}
\subsection{A Characteristic Acceleration Scale from Stellar Feedback}
\label{sec:characteristic_scale}

For historical reasons, the convention in the star formation literature has been to express Eq.~\ref{eqn:sfe} in terms of a critical surface density and it is common to assume (e.g., in the context of individual molecular clouds) that $\Sigma_{\rm tot}$ is dominated by gas. This is the case for the numerical experiments cited in \S \ref{section:derivation}. However, let us now assume that those experiments substantiate the simple force-balance argument leading to Eq.~\ref{eqn:sfe}, and that this can be generalized to a galactic context. This is not wholly certain, as these works neglect the galactic-scale context, which does imprint upon the morphologies and flow patterns of star-forming gas complexes in a manner that is more complex than in idealized setups \citep[e.g.][]{reyraposo:2015.galactic.gmcs}.
Our assumption is effectively that 
such additional effects should not alter the overall dimensional scaling in the problem, provided we recognise that the relevant acceleration should be the total gravitational acceleration including {\it all} matter, $g_{\rm tot} = G\,M_{\rm tot}/R^{2}$. 
With this definition, Eq.~\ref{eqn:sfe} can be rewritten as
\begin{align}
   \frac{M_{\ast,\,{\rm young}}}{M_{\rm gas,\,expelled}} &\sim \frac{G\,M_{\rm tot}}{\langle \dot{p}/{m}_{\ast} \rangle\,R^{2}} = \frac{g_{\rm tot}}{\gcrit}  \label{eqn:sfe.agrav}  \\ 
\nonumber \gcrit  &\equiv \langle \dot{p}/{m}_{\ast} \rangle = \pi\,G\,\Sigma_{\rm crit} \sim 4\times10^{-8}\,\rm{cm\,s^{-2}}.
\end{align}
Thus the characteristic acceleration scale inherent in stellar feedback physics, $\gcrit \sim \langle\dot{p}/{m}_{\ast} \rangle$, is on the order of the observed characteristic acceleration of disk galaxies, $\adag =1.2\times 10^{-8}\mathrm{cm\,s^{-2}}$.

Consider now what happens in different limits of the gravitational acceleration, as implied by Eq.~\ref{eqn:sfe.agrav}. If $g_{\rm tot} \ll \gcrit$, then  $M_{\ast,\,{\rm young}}/M_{\rm gas,\,expelled} \ll 1$ and the fraction of the total gas mass converted to stars is small. Thus, if dark matter initially dominates $M_{\rm tot}\left(< R\right)$, it will generally continue to do so. In the opposite limit, where $g_{\rm tot} \gg \gcrit$, the SFE is high and baryons cannot escape the galaxy. In such an instance, where feedback is ineffective, dissipative gas accretion within dark matter halos is easily sufficient to allow baryonic matter to dominate over dark matter \citep{fall.efstathiou:1980, katz:1996}. This argument implies that a constant characteristic acceleration \emph{$\gcrit \sim \adag$ should demarcate the transition between the baryon-dominated and dark matter-dominated regions of a galaxy.}

In our toy model we have applied Eq. \ref{eqn:sfe} to a single starburst event, similar to the ``monolithic collapse'' model of galaxy formation \citep{eggen62}. Strictly speaking, within the modern hierarchical paradigm of galaxy formation, this only describes phases of star formation that are ``bursty'' on the scale of the entire galaxy. However, we emphasize that the important factors here are the dimensionality and magnitude of $\gcrit$. The scaling of the SFE may therefore generalize to the more modern galaxy formation picture in which, for most galaxies, most of the stellar mass forms from gas that is continuously accreted from the intergalactic medium. 
This is predicted to be the case for practically all galaxies in the mass range observed to conform to the BTFR and RAR \citep[e.g.,][]{rodriguez.gomez:2016.illustris.insituexsitu, angles:2017.cycle}.
\vspace{-0.4cm}
\subsection{Accelerations in the Dark Matter-Dominated (``Deep-MOND'') Limit
\&\ Flat Rotation Curves}
\label{section:deepmond}
\newcommand{\gN}{g_{\rm N}}
\newcommand{\Mhalo}{M_{\rm h}}
\newcommand{\Mstar}{M_{\ast,\,{\rm young}}}

In the previous section we argued that, at large accelerations  $g_{\rm tot} \gg \adag$, the total mass is expected to be dominated by baryons and therefore $g_{\rm tot} = G\,M_{\rm tot}(<R)/R^{2} \approx g_{\rm baryon} \equiv G M_{\rm bar}(<R)/R^{2}$ (where $M_{\rm bar}(<R)$ is the baryonic mass enclosed in $R$), as is indeed observed. Consider now the low-acceleration (``deep MOND'') regime, $g_{\rm tot} \ll \adag$. In this regime, the accelerations are observed to scale approximately as $g_{\rm tot} \approx (g_{\rm baryon} \, \adag)^{1/2}$ \citep[e.g.][]{lelli:2017.rar}. Equivalently (since $g_{\rm tot} = V_{c}^{2}/R$ in terms of the circular velocity $V_{c}$), $V_{c} \approx (G M_{\rm bar}(<R)\, \adag)^{1/4}$. 
If we make the assumption that $M_{\rm bar}(<R)$ is approximately the total baryonic mass ($\equiv M_{\rm bar}$), e.g. at sufficiently large radii that the baryonic mass starts to converge, then rotation curves are asymptotically ``flat'' ($V_{c}\rightarrow$\,constant as $R\rightarrow\infty$), with a universal scale $\adag$ above which $g_{\rm tot} \approx g_{\rm baryon}$.


But in the limit $g_{\rm tot} \ll \adag$, our Eq.~\ref{eqn:sfe.agrav} implies that the stellar mass $M_{\ast} \sim (g_{\rm tot}/\gcrit)\,M_{\rm gas,\,initial} \approx (g_{\rm tot}/\gcrit) f_{\rm bar}^{0} M_{\rm tot}$, where $f_{\rm bar}^{0} \equiv M_{\rm gas,\,initial}/M_{\rm tot}$ refers to the initial ``total'' supply of baryons.\footnote{Since the SFE is low in this regime, the expelled gas mass is close to the total initial gas mass.} 
Solving for the total mass,
\begin{align}
M_{\rm tot}(<R)\approx \left( \frac{\gcrit}{g_{\rm tot}} \right) \frac{M_{*}}{f_{\rm bar}^{0}} \approx 
\left( \frac{\gcrit}{g_{\rm tot}} \right) \frac{(1-f_{\rm g})}{f^{0}_{\rm bar}} M_{\rm bar},
\end{align}
where $f_{\rm g}=M_{\rm gas}^{\rm relic}/(M_{\rm gas}^{\rm relic}+M_{*})$ is the gas fraction determined by the ``relic'' gas mass $M_{\rm gas}^{\rm relic}$ that remains after expulsion by stellar feedback. 
Using the above relations, standard Newtonian dynamics imply that the total acceleration $g_{\rm tot} = g_{\rm Newtonian}$ is:
\begin{align}
g_{\rm tot} & \approx \frac{GM_{\rm tot}(<R)}{r^{2}} \approx \left( \frac{\gcrit}{g_{\rm tot}} \right) \frac{(1-f_{\rm g})}{f_{\rm bar}^{0}} \frac{G M_{\rm bar}}{R^{2}} 
= \left( \frac{\tildegcrit}{g_{\rm tot}} \right) g_{\rm baryon}
\\
& \Rightarrow g_{\rm tot} \approx \left( g_{\rm baryon} \, \tildegcrit \right)^{1/2},
\end{align}
where $\tildegcrit \equiv \gcrit(1-f_{\rm g})/f_{\rm bar}^{0}$ only differs from $\gcrit$ at the order unity level.\footnote{At face value this predicts that the measured $g_\dagger$ in a galaxy should be $\propto (1-f_{\rm g})$. However, this is a toy derivation for a single event star formation event; a real galaxy would form over many such events, making the actual prediction non-trivial.} 
Identifying $\tildegcrit$ with  $\adag$, this may explain the observed asymptotic scaling for small accelerations.\footnote{A similar argument was given in Wheeler et al. (2019), but invoking the observed BTFR instead of the SFE scaling.}

Thus, not only the characteristic acceleration $\adag$, {\em but also} the asymptotic scalings at both low and high accelerations that define the observed RAR (and MOND, by construction) can potentially be explained as consequences of how stellar feedback acts.  
Equivalently, in this picture, the fact that observed rotation curves are approximately flat at large radii does not require a ``conspiracy'' or ``coincidence'' between baryons and dark matter: rather, stellar feedback ensures the baryons self-regulate with the scaling needed to ensure approximately flat $V_{c}$.\footnote{We note the factor $(1-f_{\rm g})/f_{\rm bar}^{0}$ is not exactly constant at large radii (though for a galaxy in an NFW dark matter halo it varies extremely weakly with radius): this may reflect the fact that observed rotation curves are not perfectly flat \citep[e.g.,][]{salucci:2000.rotation.curves,courteau:2015.rotation.curves}.}

\vspace{-0.5cm}

\subsection{Expressing $\adag$ in Fundamental Constants}

According to the picture proposed here, $\adag \sim \gcrit \sim 0.3\,\ffb$. But $\ffb$ itself can be understood in terms of the stellar IMF and the physics of massive stars. 
As discussed in \S \ref{section:derivation}, the stellar population-averaged $\dot{p} \sim L/c$ (i.e.\ $\ffb \sim c^{-1}\,(L/M)_{\rm \ast,\,young}$) for a variety of feedback mechanisms, and the most massive stars dominate the luminosity and feedback. The luminosities of these stars are set by the Eddington limit:
\begin{equation}
L_{\rm Edd,\,i} = \frac{4\pi G m_{\rm p} c}{\sigma_{\rm T}} M_{i},
\end{equation}
where $M_{i}$ is the mass of an individual massive star and $\sigma_{\rm T}$ is the Thomson cross-section. So $(L/M)_{\rm \ast,\,young} \sim \fmassive\,L_{\rm Edd,\,i}/M_{i}$ where the $\fmassive$ is the mass fraction in massive stars: to reproduce the more accurate full stellar population calculation for a standard IMF \citep{starburst99}, $\fmassive\sim 0.03$.\footnote{\citet{krumholz:2011.imf} argue that the form of the IMF, and by extension $\fmassive$, are also expressible in terms of fundamental constants. 
} Thus we have $\ffb \sim f_{\rm massive}\,(4\pi\,G\,m_{\rm p})/\sigma_{\rm T}$. Putting these together, we obtain:
\begin{align}
    \adag &\sim \gcrit \sim (4\,\fmassive)\,\frac{G m_{\rm p}}{\sigma_{\rm T}} \sim 0.1\,\frac{G m_{\rm p}}{\sigma_{\rm T}} \sim 0.5\,G m_{\rm p} \left(\frac{m_{\rm e} c}{\alpha h} \right)^{2},
\end{align}
where the last expression uses $\sigma_{\rm T} \equiv \frac{8 \pi}{3} \left( \frac{\alpha \hbar c}{m_{\rm e} c^{2}} \right)^{2}$, in terms of the fine-structure constant $\alpha$, the Planck constant $\hbar=h/2\pi$, the electron mass $m_{e}$, and the speed of light $c$.

\vspace{-0.5cm}

\section{Discussion}

\subsection{Departures from universality}
Recent observational analyses have cast serious doubt upon a {\it fundamental} RAR to which all galaxies must conform exactly \citep{rodrigues:2018.no.fundamental.acceleration, stone:2019.rar.scatter, chang:2019.rar.scatter}, contrary to previous claims that the observed scatter is fully consistent with measurement errors \citep{li:2018.rarfits}. Hence, whatever the origin of $\adag$, it is likely emergent rather than fundamental in nature. This fits with the picture presented in this paper: although $\gcrit$ sets a characteristic scale, there is no reason why galaxies should conform {\em exactly} to a single RAR. For example, galaxy formation is the product of both in-situ star formation and hierarchical merging, with merging becoming increasingly important at higher masses, e.g. for massive elliptical galaxies \citep[e.g.,][]{angles:2017.cycle}. 

Feedback from active galactic nuclei (AGN), rather than stars, is also expected to be most important in massive galaxies.  Interestingly, while lower-mass disk galaxies prefer the characteristic acceleration described here, the massive ellipticals do not (though they do preserve ``memory'' of the mass profiles of merged galaxies; see \citealt{boylan-kolchin:2005}). Thus, effects such as variations in galaxy assembly history and AGN feedback should cause galaxies to deviate from the RAR and BTFR.
\vspace{-0.4cm}

\subsection{Effect of IMF variations}
Another source of scatter would be variations in the IMF: $\dot{p}/m_\ast$ is sensitive to the massive stellar content of a stellar population, so within a given galaxy the observed $\adag$ could vary with it accordingly. 

{\it Random} variations in $f_{\rm massive}$ are present in any stellar population ``sampled" from an IMF due to the finite number of stars, but in particular the effect of ``incomplete" sampling of the IMF upon $\dot{p}/m_\ast$ becomes pronounced for stellar populations less massive than $10^4 \rm~M_\odot$ \citep{murray:2010.wmap, kim:2016.dusty.hii}. We thus expect increased scatter in the BTFR and RAR in ultra-faint dwarf (UFD) galaxies with $M_\ast < 10^4~\rm M_\odot$.

It is also possible for the IMF to vary systematically from one galaxy to another, and for the inferred $\adag$ to vary with it. However, the factor $f_{\rm massive}$ is fairly well-constrained in an average sense: if it varied strongly with galaxy or local environmental properties, the slopes and normalizations of the Schmidt-Kennicutt 
and $M_{\ast}-M_{\rm halo}$ relations (which are largely set by feedback in current theories) would be quite different. Furthermore, direct observations are consistent with an IMF that is common to all galaxies \citep{bastian:2010.imf.universality, offner:2014.imf.review}. Therefore, while variations in the measured $\adag$ could conceivably be driven by IMF variations, it seems likely that they would be dominated by other sources of variations, such as galactic environment and assembly history.

Insofar as the IMF \textit{is} roughly universal across cosmic time and feedback does not depend strongly on factors such as metallicity, the arguments presented above imply that (1) the characteristic acceleration scale should not vary systematically with redshift and (2) the observation that $\adag \approx c\,H_0$ in the low-redshift Universe \citep[][]{mond} is a numerical coincidence.
\vspace{-0.4cm}
\subsection{Additional Details from Cosmological Calculations}
\label{sec:cosmo_details}
In $\Lambda$CDM, dark-matter-only simulations predict NFW-like dark matter profiles with $\rho \propto r^{-1}$ on small scales.
In such halos, the maximum (central) acceleration is nearly constant, $g_{\rm tot}^{\rm cen} \sim 3\times10^{-9}~\rm{cm\,s^{-2}}$ (or central $\Sigma_{\rm DM}^{\rm cen} \sim 100\,\rm{M_{\odot}\,pc^{-2}} < \Sigma_{\rm crit}$), with very weak dependence on halo mass and/or redshift,  
although the presence of baryons and stellar feedback can strongly alter this  \citep[e.g.][]{navarro.1996.cuspcore, governato:2010.cuspcore, onorbe:2015.cusp.core}. 
A ``pure'' dark halo therefore has $g_{\rm tot} \lesssim \gcrit$ at all radii.\footnote{This alone can explain, in part, why systems with $g_{\rm tot} \gg \adag$ must be baryon-dominated, but it does not explain why systems with $g_{\rm tot} \ll \adag$ could not also be baryon-dominated.}  But if all of the available baryons fall in, conservation of specific angular momentum implies circularization in a disk with extent $r \sim \lambda\,R_{\rm vir}$ \citep[where $R_{\rm vir}$ is the virial radius; e.g.,][]{fall.efstathiou:1980, mmw}. The acceleration in the galaxy $g_{\rm tot}\sim 2\times 10^{-8}\,(1+z)^{2}\,\rm{cm\,s^{-2}}$ ($\Sigma_{\rm bar}\sim 500\,(1+z)^{2}\,\rm{M_{\odot}\,pc^{-2}}$) then approaches the critical value. If the baryons lose additional angular momentum, or fall in at high redshift, then the center can become arbitrarily baryon-dominated in principle.


Properly accounting for the complexities of galaxy formation 
requires much more detailed modeling. 
But of course, this is what cosmological simulations and semi-analytic models do. These calculations have indeed shown 
that the characteristic acceleration scale $\adag$, and more detailed scaling relations such as the RAR and baryonic Tully-Fisher relation (BTFR), emerge naturally in $\Lambda$CDM {\em provided} the models accurately treat 
stellar feedback (see references in \S 1). %
More specifically, previous studies have shown that in $\Lambda$CDM models that produce the ``correct'' galaxy sizes and masses, the characteristic $\adag$ appears in rotation curves as observed; \citet{wheeler:2018.rar} showed more explicitly that as long as galaxies are broadly consistent with the observed BTFR, this is essentially guaranteed. And it is well known that stellar feedback with $\langle \dot{p}/m_{\ast} \rangle$ similar to the values assumed here, based on standard stellar evolution models, is necessary to reproduce these scaling relations in $\Lambda$CDM \citep[e.g.,][]{somerville:review}. 
The simple arguments presented in this paper do not supplant these much more detailed and sophisticated calculations. Rather, they help demonstrate that the observed universal acceleration scale of galaxies does not require a ``conspiracy'' between many different fine-tuned components nor a modified theory of gravity:  {\em stellar feedback contains a characteristic acceleration} that maps directly to the acceleration scale seen in galaxies. 
\vspace{-0.5cm}
\section*{Acknowledgements}
We thank James Bullock, Manoj Kaplinghat, and Jonathan Stern for useful discussions, and the anonymous referees for various comments that improved and clarified this work. MYG is supported by  the  CIERA  Postdoctoral  Fellowship  Program. 
MBK acknowledges support  from  NSF  grants  AST-1517226, AST-1910346  and  CAREER  award  AST-1752913, NASA grant NNX17AG29G and grants HST-AR-13888, HST-AR-13896, HST-AR-14282, HST-AR-14554, HST-AR-15006, HST-GO-12914, and HST-GO-14191 from the Space Telescope Science Institute, which is  operated  by  AURA,  Inc.,  under  NASA  contract  NAS5-26555.
CAFG is supported by NSF through grants AST-1517491, AST-1715216, and CAREER award AST-1652522, by NASA through grant 17-ATP17-0067, and by a Cottrell Scholar Award from the Research Corporation for Science Advancement. Support  for  PFH   was  provided by  an  Alfred  P.  Sloan  Research  Fellowship,  NSF  Collaborative Research  Grant  \#1715847  and  CAREER  grant  \#1455342,  and NASA grants NNX15AT06G, JPL 1589742, and 17-ATP17-0214. 
\vspace{-0.5cm}

\bibliographystyle{mnras}
\bibliography{master} 






\bsp	
\label{lastpage}
\end{document}